\documentclass{article} 
\usepackage[dvips]{graphicx,epsfig,color} \usepackage{wrapfig,rotating} 
\usepackage{amssymb,amsmath,array}
\usepackage{glas}
\begin{document}
\begin{titlepage} {GLAS-PPE/2009-05} {18$^{\underline{\rm{th}}}$ June 
2009}
\title{
Inclusive \boldmath$K^0_SK^0_S$ Resonance Production\\
in \boldmath$ep$ Collisions}   

\author{D.H. Saxon  (on behalf of the ZEUS Collaboration)
%
\thanks{Supported by UK Science and Technology Facilities Council}
%
\vspace{.3cm}\\
%
%
}


\begin{abstract}
Resonant structure in the inclusive $K^0_s K^0_s$ mass spectrum is 
interpreted via interference between three tensor mesons plus the 
production of a glueball candidate state.

\vspace{0.5cm}
\begin{center}
{\em DIS 2009 conference}\\
{\em Madrid, April 26-30, 2009}
\end{center}
 \end{abstract}
\newpage
\end{titlepage}
\section{Introduction and data set}
  We report on the $K^0_s K^0_s$ mass spectrum\cite{url,zeus},
seen  
using the 
full HERA 
data set ($0.5{\rm pb}^{-1}$, 77\% from HERA-II) with the ZEUS 
detector.
The data sample of 672418 $K^0_s K^0_s$ pairs is 90\% from 
photoproduction.

Recall that the $J^P = 1^-$ $\phi$-meson decays to $K^0 \overline {K}^0$ 
as $K^0_S K^0_L$ but never to 
$K^0_s K^0_s$ or to $K^0_L K^0_L$. This is an example of the non-local 
nature of quantum mechanics. The Bose symmetry of the $K^0_s K^0_s$ pair 
forces $J^P = 0^+, 2^+, 4^+ $ etc. making this a good place to search
for glueballs.

\begin{figure} [htb]
\centerline{\includegraphics[width=0.45\columnwidth]{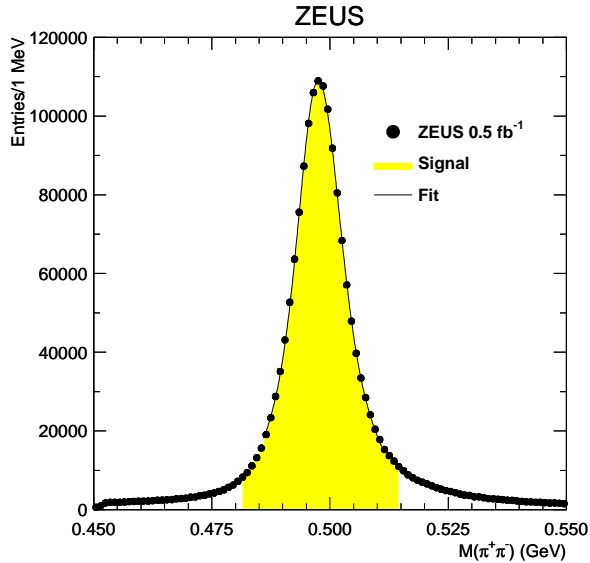}}
\caption{Measured $\pi^+ \pi^-$ mass spectrum.}
\label{saxon_davidKK.fig1.eps}
\end{figure}

 Quenched-approximation lattice gauge calculations \cite{lqcd} suggest 
the lightest glueball states are: $J^{PC}=0^{++}$ at $1710 \pm 50 \pm 80 
~\rm{MeV}$ and $J^{PC}= 2^{++}$ at $2390 \pm 30 \pm 120~ \rm{MeV}$.
Four states are found with $J^{PC}=0^{++}$: $f_0(980)$, $f_0(1370)$,
$f_0(1500)$ and $f_0(1710)$, consistent with three $q \overline{q}$ 
states and 
one $gg$ state. The physical states can be mixtures of these.
$q \overline {q}$ states are produced as leading hadrons in direct 
photoproduction or in fragmentation. $gg$ states can be produced as 
leading hadrons 
in resolved photoproduction or in fragmentation.  

\begin{figure}[htb]
\centerline{\includegraphics[width=0.90\columnwidth]{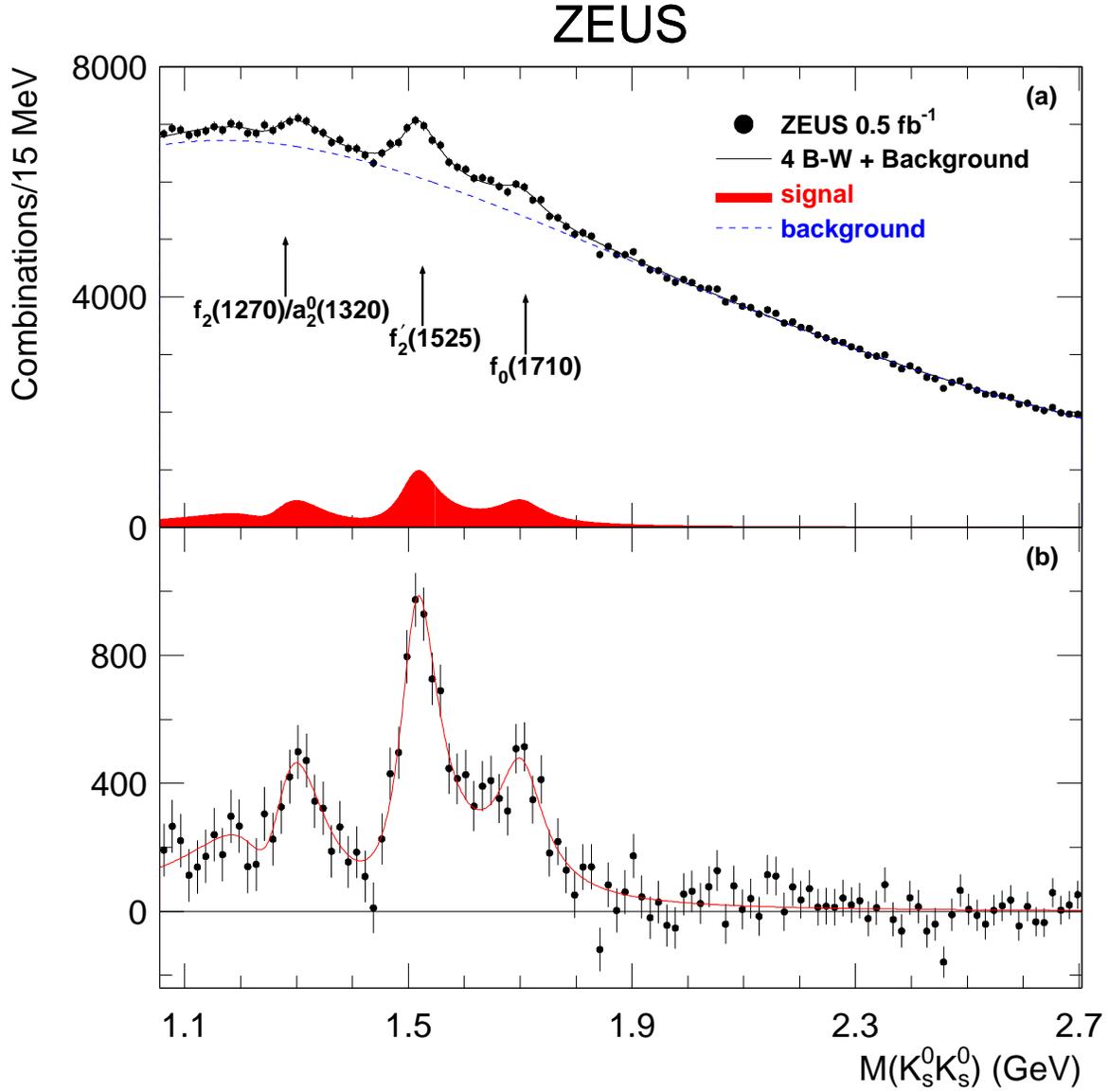}}
\caption{$K^0_sK^0_s$-mass (a) data and coherent fit (b) same with smooth 
background 
subtracted.}
\label{saxon_davidKK.fig2.eps}
\end{figure}

The data were produced in the ZEUS detector,
 relying most on the central 
tracking (72 layers, B = 1.4T, $\sigma \simeq 160 \mu\rm{m}$) and the 
microvertex detector ($\sigma \simeq 25-35 \mu\rm{m}$) for reconstruction 
of 
the 
$K^0_s \rightarrow \pi^+ \pi^-$ decay.
  Both tracks from the same secondary decay 
vertex were assumed to be
charged pions and the invariant mass, $M(\pi^+\pi^-)$, of each track
pair was calculated. The $K^0_s$ candidates were selected by requiring:
$M(e^{+}e^{-})\ge 50~ \rm{MeV}$, where the electron mass was assigned
to each track, to eliminate tracks from photon conversions and
$M(p\pi)\ge 1121~\rm{MeV}$, where the proton mass was assigned to the 
track
  with higher momentum, to eliminate $\Lambda$ and
 $\overline{\Lambda}$ contamination to the $K^0_s$ signal.

We require
$p_T ( K_S^0 )\ge 0.25\rm{GeV}$ and $|\eta ( K_S^0 ) |\le 1.6$;
$\theta_{2D} <~0.12$ ($\theta_{3D} <~0.24$), where $\theta_{2D}$
($\theta_{3D}$) is the two (three) dimensional collinearity angle between 
the $K^0_s$
momentum vector and the vector defined by the interaction point and the
vertex. (For $\theta_{2D}$, the $XY$ plane was used.)
The cuts on the collinearity
angles significantly reduced the non-$K^0_s$ background
 in the data during the 2004--2007 period, using microvertex detector 
information.
After all these cuts, the decay length distribution of the resulting 
$K^0_s$
candidates peaks
at $\approx 2\rm{cm}$.

Events with at least two $K^0_s$ candidates
in the mass range of $481\le M(\pi^+\pi^-)\le 515\rm{MeV}$
 were accepted for further 
analysis.
Figure~\ref{saxon_davidKK.fig1.eps} shows the 
$M(\pi^+\pi^-)$ distribution 
of these
$K^0_s$ candidates.  
Figure~\ref{saxon_davidKK.fig2.eps} shows the $M(K^0_s K^0_s)$ 
distribution. The $K^0_sK^0_s$ mass resolution is typically 12 MeV.

\begin{figure}[htb]
\centerline{\includegraphics[angle=-90,width=0.45\columnwidth]{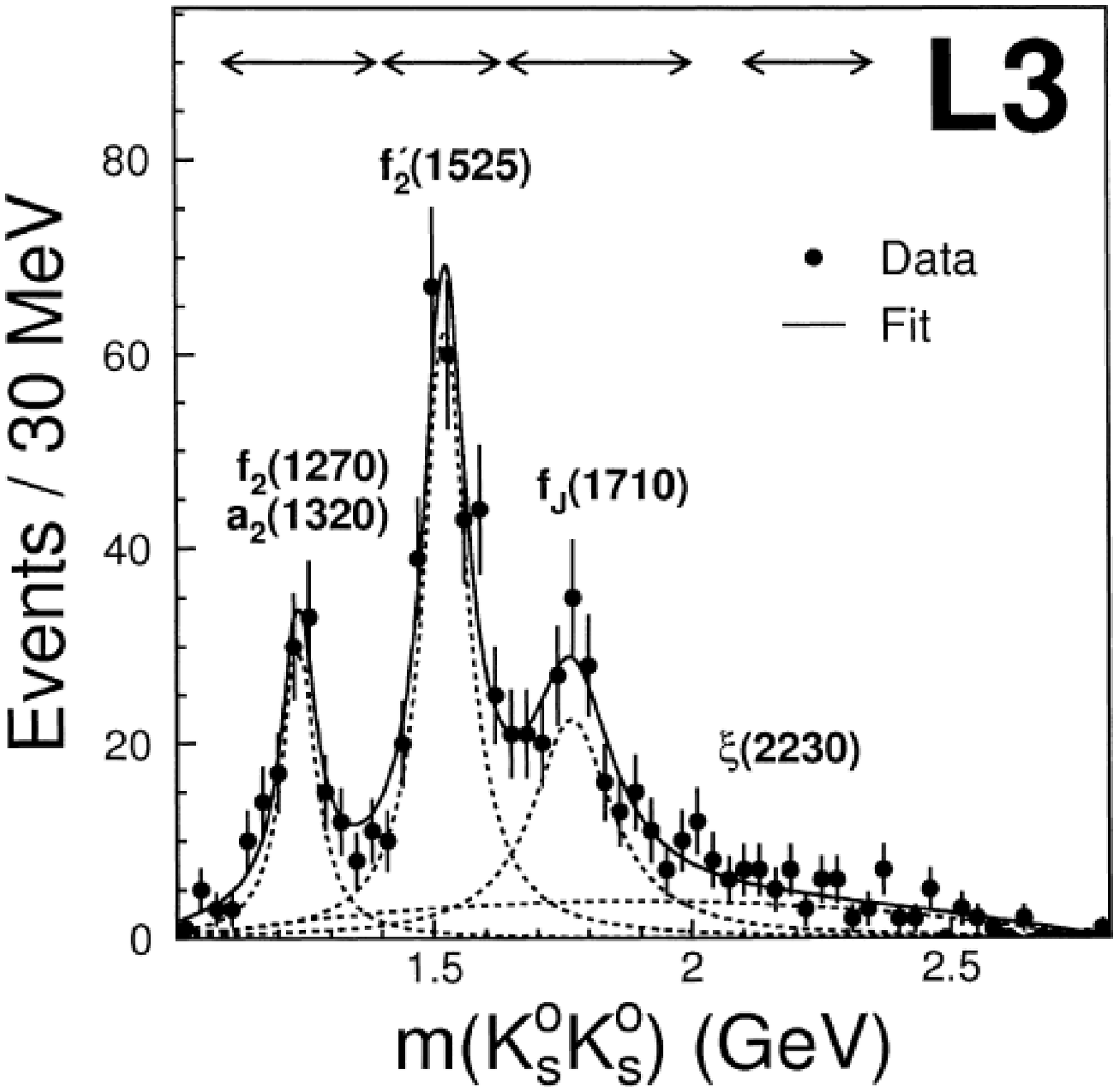}}
\caption{$\gamma \gamma \rightarrow K^0_s K^0_s$ results.}
\label{saxon_davidKK.fig3.eps}
\end{figure}

\section{Interpretation}

The first $M(K^0_sK^0_s)$ fit (not shown) used a smooth background plus 
three 
incoherent 
Breit-Wigners. A health warning is in order: `Breit-Wigner plus 
background' fits have strong correlations between the fitted BW intensity, 
the width and the background. There is a long history!

The $\chi^2 /NDF$ is good (96/95) but the fit is poor near 1300 MeV and 
the width of the bump in the $f_2(1270)/a^0_2(1320)$ region is $61 \pm 11$ 
MeV, far too narrow for the $f_2$ and $a^0_2$ for which the PDG values are 
$176 \pm 17$ and $114 \pm 14$ MeV.

A similar result was obtained for the exclusive process $\gamma \gamma 
\rightarrow K^0_s K^0_s$ by the L3 collaboration \cite{L3}. The fitted 
mass spectrum is shown in Figure 3. Their fitted $f_2/a^0_2$ peak has a 
mass and width of $1239 \pm 6$ and $78 \pm 19$ MeV.

\begin{figure}[htb]
\centerline{\includegraphics[width=0.75\columnwidth]{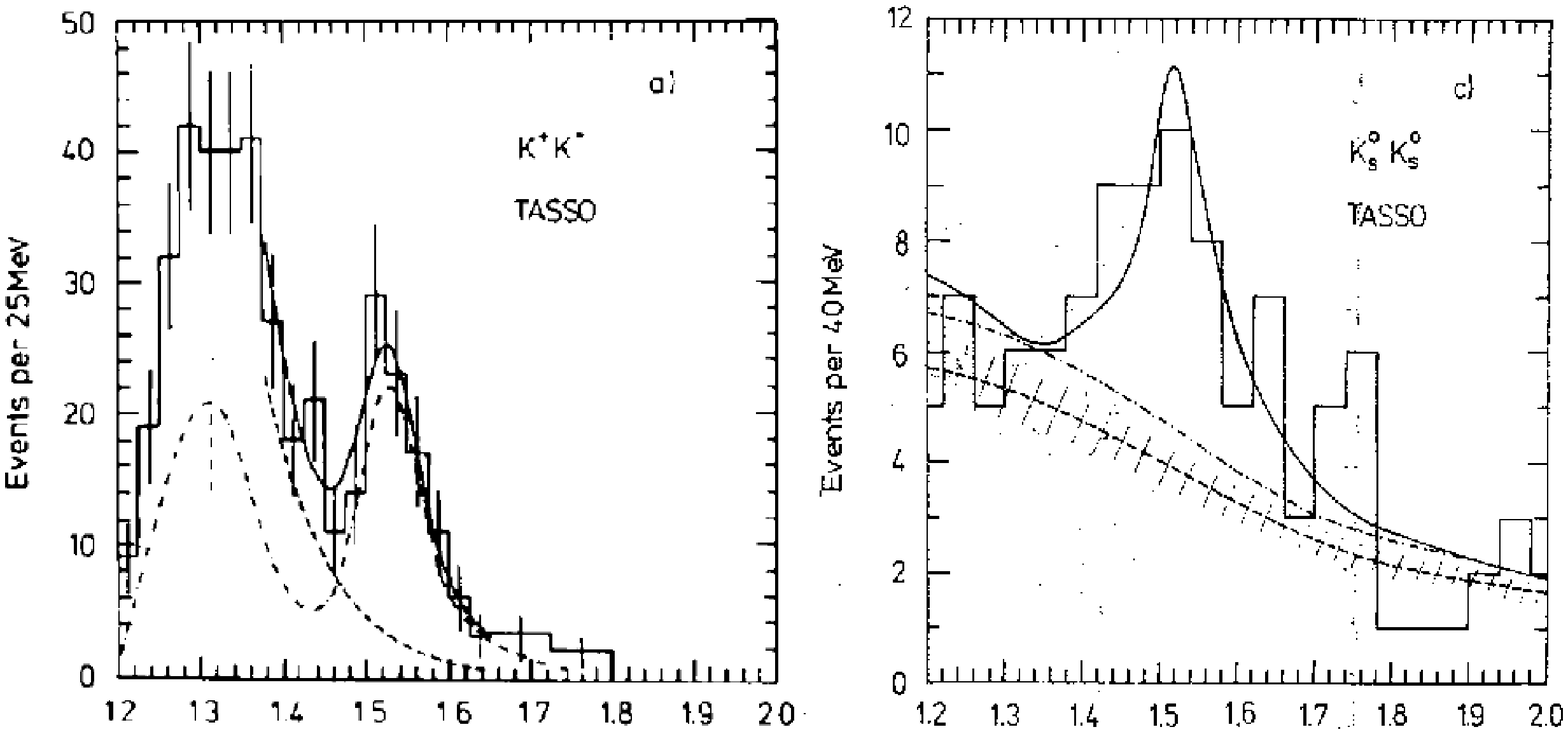}}
\caption{$\gamma \gamma \rightarrow K^+K^-$ and $K^0_sK^0_s$ results.}
\label{saxon_davidKK.fig4.eps}
\end{figure}

The TASSO collaboration measured the same process \cite{TASSO} 
and also measured $\gamma \gamma \rightarrow K^+K^-$. The $K^0_s K^0_s$ 
spectrum again shows an $f'_2(1525)$ signal but no trace of any 
enhancement around 1300 MeV. The $K^+ K^-$ result also shows a clear 
signal around 1525 MeV but has a major and broad enhancement in the 
$f_2/a^0_2$ 
region around 1300 MeV.
 (see Fig. 4).

Health warning 2: The $f_2 (1270), a_2 (1320)$ and $f'_2 (1525)$ all have 
$J^p = 2^+$. In exclusive $\gamma \gamma$ production these must 
therefore interfere 
in the mass 
spectrum. Time reversal invariance makes the coefficients of their 
production amplitudes real.

The reaction $\gamma \gamma \rightarrow KK$ proceeds via electromagnetic 
coupling to the quark charges. Faiman, Lipkin and Rubinstein \cite{Faiman} 
use the 
quark 
structure of the resonant states. Thus for the $I=0$ $f_2(1270)$
 the quark 
content is $(u \overline{u} + d \overline{d})/\sqrt{2}$ giving a charge 
amplitude ratio factor $(2/3 \times 2/3 + 1/3 \times 1/3)/2 = 5/18$. For 
the $I=0$ 
$f_2(1525)$ the content $s \overline{s}$ gives a factor 2/18, and the
$I=1$ $a^0_2(1320)$
 the quark 
content is $(u \overline{u} - d \overline{d})/\sqrt{2}$ giving a charge 
amplitude ratio factor $(2/3 \times 2/3 - 1/3 \times 1/3)/2 = \pm{3/18}$, 
where the + sign applies to the $K^+K^-$ final state and the - sign to 
$K^0_s K^0_s$. Since the $f_2$ and $a_2$ are so close in mass we expect 
predominantly constructive interference between them in $K^+K^-$ and 
predominantly destructive interference in $K^0_s K^0_s$, as observed by
TASSO and L3.

The form used in the coherent fit to the ZEUS $ep$ inclusive $K^0_sK^0_s$
 mass spectrum, $f(m)$, is
\begin{centering}
\vspace{3mm} \\
$f(m) = a \times| 5 B_{f(1270)}(m) - 3 B_{a(1320)}(m) + 2 B_{f(1525)}(m) 
|^2 + b \times
|B_{f(1710)}(M)|^2 + c \times U(m)$ 
\vspace{3mm} \\
\end{centering}
\noindent
where $B_M(m)$ is the relativistic Breit-Wigner,
$B_M(m) = M \sqrt{\Gamma}/(M^2 - m^2 - iM \Gamma)$, and $U(m)$ is a smooth 
background function.

\section{Results}
Figure 1 shows the resulting coherent fit, with the fitted background 
subtracted in 
figure 1(b). Compared to the no-interference fit the $\chi^2/NDF$ improves 
to 86/97, which can be viewed as a $3 \sigma$ improvement. Note that in 
this fit the $J^P = 2^+$ states couple directly to the exchanged photon.

Fits without the $f(1710)$ are strongly disfavoured, with $\chi^2/NDF = 
162/97$.

Table 1 compares the fitted parameters to Particle Data Group values. 
Mostly they are in good agreement. All the widths agree but the $a_2$ mass 
is still low. Figure 5 compares the $f'_2(1525)$ and $f_0(1710)$ results 
to previous measurements.

\begin{table}
\centerline{\begin{tabular}{|l|l|l|l|l|}
\hline
State       & $M$(fit)&  $\Gamma$(fit) & $M$(PDG) & $\Gamma$(PDG) 
\\\hline
$f_2(1270)$ & $1268 \pm 10$ & $176\pm 17$ & $1275.4 \pm 1.1 $& $185 \pm 
3$ 
\\\hline
$a^0_2(1320)$& $1257 \pm 9$& $114 \pm 14$& $1318.3\pm0.6$ &$ 107 \pm 5$ 
\\\hline
$f'_2(1525)$&$1512 \pm3^{+1.4}_{-0.5}$&$83 \pm 9^{+5}_{-4}$& $1525 \pm 
5$& $73 ^{+6}_{-5}$ \\\hline
$f_0(1710)$&$1701 \pm 5 ^{+9}_{-2}$& $100 \pm 24 ^{+7}_{-22}$& $ 1724 \pm 
7$& $137 \pm 8$ \\
\hline
\end{tabular}}
\caption{Coherent fit: camparison to PDG.}
\label{tab:fit}
\end{table}

\begin{figure}[htbp]
$\begin{array}{cc}
&
\includegraphics[width=0.50\columnwidth]{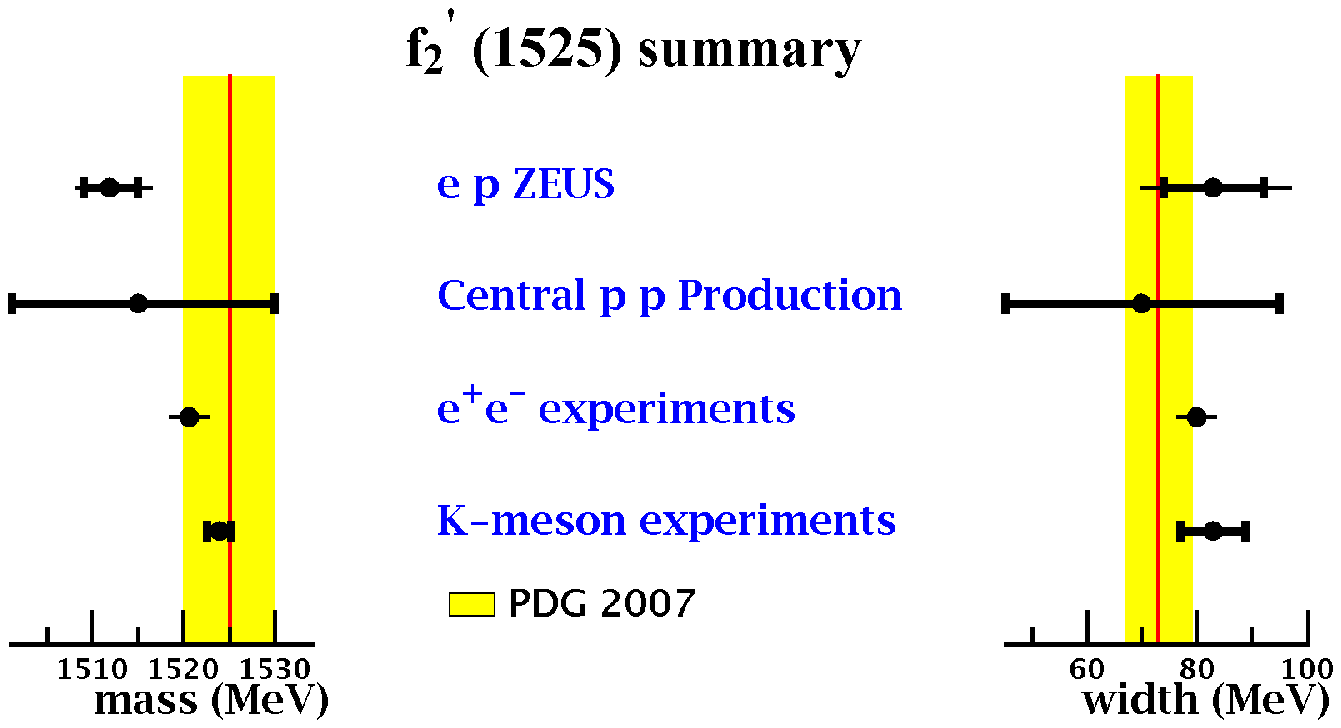}
\includegraphics[width=0.50\columnwidth]{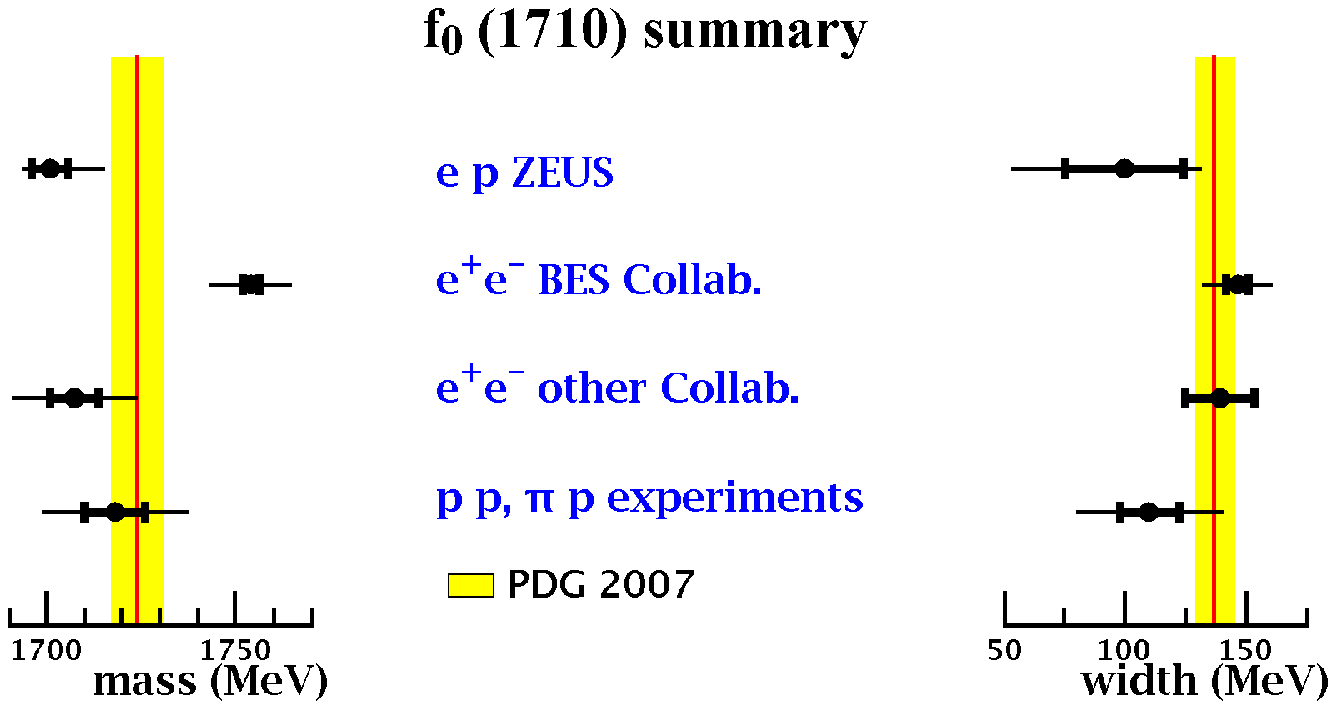}
\\
\end{array}$
\caption{Comparisons to previous measurements}
\label{saxon_davidKK.fig5b.ps}
\end{figure}

In conclusion, ZEUS have made a high-statistics study of the $K^0_s K^0_s$ 
system. Only $J^P = \rm{even}^+$ states are possible.
There is evidence for the coherent production of three $J^{PC} = 2^{++}$
states. Negative $f_2/a_2$ interference suggests coupling to the exchanged 
photon.

Production of the $f_0(1710)$ is clearly observed. This cannot be a pure 
glueball if it is the same state as the $f_J(1710)$ seen in $\gamma 
\gamma$ collisions.

\begin{footnotesize}



%

\end{footnotesize}


\end{document}